\documentclass[apj]{emulateapj}

\usepackage{bm}
\usepackage{graphicx}
\usepackage{epsf}
\usepackage{graphics}
\usepackage{amsmath}

\def\mpc{\,h^{-1}{\rm Mpc}}
\def\kpc{\,h^{-1}{\rm kpc}}

\def\msun{\,h^{-1}{\rm M}_\odot}
\def\bx{{\boldsymbol x}}
\def\bt{{\boldsymbol t}}
\def\cluster{{\tt cluster$\;$}}
\def\sheet{{\tt sheet$\;$}}
\def\filament{{\tt filament$\;$}}
\def\void{{\tt void$\;$}}

\usepackage{color}

\makeatletter

\newcommand{\Rmnum}[1]{\expandafter\@slowromancap\romannumeral #1@}
\makeatother

\shorttitle{galaxies in the filaments}
\shortauthors{Zhang et al.}

\begin{document}


\title{Alignments of galaxies within cosmic filaments from SDSS DR7}

\author{Youcai Zhang\altaffilmark{1}, Xiaohu Yang\altaffilmark{1,2}, Huiyuan
  Wang\altaffilmark{3,4}, Lei Wang\altaffilmark{5}, H. J. Mo\altaffilmark{4},
  Frank C. van den Bosch\altaffilmark{6}}

\altaffiltext{1}{Key Laboratory for Research in Galaxies and Cosmology,
  Shanghai Astronomical Observatory; Nandan Road 80, Shanghai 200030,
  China; E-mail: yczhang@shao.ac.cn}

\altaffiltext{2}{Center for Astronomy and Astrophysics, Shanghai Jiao
Tong University, Shanghai 200240, China; Email: xyang@sjtu.edu.cn}

\altaffiltext{3}{Key Laboratory for Research in Galaxies and Cosmology,
University of Science and Technology of China, Hefei, Anhui 230026, China}

\altaffiltext{4}{Department of Astronomy, University of Massachusetts,
Amherst MA 01003-9305, USA}

\altaffiltext{5}{Purple Mountain Observatory, the Partner Group of MPI f\"{u}r
  Astronomie, 2 West Beijing Road, Nanjing 210008, China}

\altaffiltext{6}{Department of Astronomy, Yale University, P.O. Box 208101,
  New Haven, CT 06520-8101, USA}

\begin{abstract}
  Using a sample of galaxy groups selected from the Sloan Digital Sky Survey
  Data Release 7 (SDSS DR7), we examine the alignment between the orientation
  of galaxies and their surrounding large scale structure in the context of
  the cosmic web. The latter is quantified using the large-scale tidal field,
  reconstructed from the data using galaxy groups above a certain mass
  threshold. We find that the major axes of galaxies in filaments tend to be
  preferentially aligned with the directions of the filaments, while galaxies
  in sheets have their major axes preferentially aligned parallel to the plane
  of the sheets. The strength of this alignment signal is strongest for red,
  central galaxies, and in good agreement with that of dark matter halos in
  N-body simulations. This suggests that red, central galaxies are well
  aligned with their host halos, in quantitative agreement with previous
  studies based on the spatial distribution of satellite galaxies. There is a
  luminosity and mass dependence that brighter and more massive galaxies in
  filaments and sheets have stronger alignment signals.  We also find that the
  orientation of galaxies is aligned with the eigenvector associated with the
  smallest eigenvalue of the tidal tensor.  These observational results
  indicate that galaxy formation is affected by large-scale environments, and
  strongly suggests that galaxies are aligned with each other over scales
  comparable to those of sheets and filaments in the cosmic web.
\end{abstract}


\section{Introduction}\label{sec_intro}

Numerical simulations and large galaxy redshift surveys have revealed a
striking cosmic web, composed of clusters, filaments, sheets and voids as the
primary building blocks of the large scale structure \citep[see][and
  references therein]{Bond1996}. According to the standard paradigm for galaxy
formation, these structure arise from gravitational instability amplifying
primordial, gaussian density fluctuations in a nearly homogeneous and
isotropic early Universe.  Galaxies are the visible component of
gravitationally bound systems dominated by dark matter halos, which form via
gravitional collapse of density perturbations that become non-linear
\citep[e.g.,][]{MBW10}. Since dark matter halos grow in mass by accretion, and
are subjected to the tidal forces from their surroundings, the structure and
formation of both dark matter halos and galaxies is affected by their
large-scale environment \citep[e.g.,][]{White1984, Byrd1990, Gott2001,
  Arag2012, Muld2012}.

From a dynamical point of view, matter tends to flow out of the voids, then
accretes onto the sheets, collapses to the filaments, and finally accumulates
onto large clusters at the intersection of multiple filaments. This accretion
history is expected to impact the properties of halos and galaxies formed in
them \citep[e.g.][]{Zhao2003, Zhao2009}, as well as their clustering strength
\citep{Sheth2004, Gao2005, Jing2007, Wang2007, Dal2008, Hahn2009}. Indeed,
numerous studies have found, using numerical simulations, that the spins
and/or shapes of dark matter halos are aligned with their large-scale
structure \citep{Fal2002, Col2005, Kasun2005, Bas2006, Bett2007, Cue2008,
  Paz2008, Zhang2009, Wang2011, Cod2012, Trow2013}.  In particular, the major
axes of halos are found to lie preferentially on the shells of voids
\citep{Pati2006, Brun2007, Cue2008}, in the plane of sheets \citep{Arag2007,
  Hahn2007b, Zhang2009}, and along the direction of filaments
\citep{Altay2006, Arag2007, Hahn2007a, Zhang2009, Lib2013}.  Since galaxy
properties, such as spin and shape, are expected to be affected by their host
halos \citep{Ber2008, Deb2013}, alignments between galaxies and large-scale
environment are also expected in the current paradigm of galaxy formation
\citep{Jim2010}.

Observationally, numerous studies have characterized various forms of
alignment between the orientation of galaxies and their large scale
environment. These include the alignment between the orientation of central
galaxies and the spatial distribution of their satellite galaxies, or the
major axis of its host group \citep{Sales2004, Bra2005, Yang2006, Azz2007,
  Fal2007, WangY2008, WangY2010, Agu2010, Nie2011, LiZ2013}, the alignment
between the orientation of satellite galaxies and central-satellite position
vector \citep{Per2005, Fal2007, Hao2011, Sch2013}, and the alignment between
galaxy group shape and the large scale structure \citep{Fal2009, WangY2009,
  God2010, God2011, God2012, Paz2011, Wang2011, Ski2012, LiC2013}.  All these
observational studies reveal a strong alignment between galaxies and their
surrounding large-scale structure. Moreover, the strength of the alignment is
found to depend on a variety of galaxy properties. Typically, the alignment
effects are found to be much stronger for red galaxies than for blue galaxies,
and to increase with the mass of the galaxy or its host halo.  For example,
using a large galaxy redshift catalog from SDSS DR6, \citet{Fal2009} revealed
an overabundance of reference galaxies along the major axes of red galaxies
out to projected distances of $60\mpc$, but no alignment signal was detected
for blue galaxies. Using the CMASS galaxy sample at $z \sim 0.6$,
\citet{LiC2013} found that more massive galaxies are more strongly aligned
with the large-scale structure up to scales of $70\mpc$. Using galaxies in the
SDSS, \citet{Tempel2013b} and \citet{Tempel2013c} found that the spin axis of
bright spiral galaxies aligns with the host filament but with significant
correlation with the sheet normal.

Theoretically, alignment studies with $N$-body simulations have so far focused
on statistics on the orientations of halos with respect to directions defined
by cosmic web, such as filaments, sheets or voids \citep{Brun2007, Zhang2009,
  Wang2011}. The alignment signals found here are in general much stronger
than those obtained in observations for galaxies. Such kind of difference
indicates that, due to complex accretion and merger histories, there is a
misalignment between galaxies and their host halos.  Indeed, \citet{Fal2009}
found a mean projected misalignment between a halo and its central region of
$\sim 25^\circ$.  \citet{WangY2008} suggested the misalignment followed a
Gaussian distribution with a mean of $0^\circ$ and a dispersion of $\sim
23^\circ$ \citep[see also][]{Oku2009, LiC2013}.  Using hydrodynamical
simulations, \citet{Hahn2010} found that galaxy spin is strongly aligned with
the inner dark matter halo regions where the discs reside, but the alignment
of galaxy spin with the spin of the entire host halo is much
weaker. Nevertheless, massive galaxy discs do have spins preferentially
aligned with their host filaments, while the spin of discs in lower mass halos
shows alignment with the intermediate principal axis of the large-scale tidal
field \citep{Hahn2010}.

In this paper, we study the alignments between galaxy shapes and the
surrounding large-scale environments, such as filaments and sheets. Different
from earlier investigations, here we use a large galaxy group catalog
constructed from the Sloan Digital Sky Survey Data Release 7 \citep[SDSS
  DR7,][]{Aba2009} so that we can distinguish between central and satellite
galaxies.  Furthermore, we use the cosmic tidal field in the SDSS DR7 survey
volume directly reconstructed from the group catalog, using the method of
\citet{Wang2012} that has been shown to be reliable.  Based on the signs of
the eigenvalues of this tidal field we classify the cosmic web into four
categories: clusters, filaments, sheets, and voids.  We then measure the
angles between the projected orientations of the major axes of galaxies and
the eigenvectors of the local tidal tensor, and study how these alignments
depend on galaxy properties.  In order to interpret our results, we compare
our findings with the (projected) alignments of dark matter halos and their
cosmic web environment obtained from a large $N$-body simulation.

This paper is organized as follows. In Section \ref{sec_data}, we present the
data and the methodology used to measure the various alignment signals, both
for galaxies in the SDSS DR7 and for dark matter halos in our $N$-body
simulation. The alignments of galaxies and dark matter halos with respect to
the cosmic web are presented in Section \ref{sec_result}, and we summarize and
discuss our results in Section \ref{sec_summary}.

\section{Data and Methodology}
\label{sec_data}

In this section we outline how we characterize the cosmic web using galaxy
groups identified in the SDSS DR7 galaxy catalog, and we describe the method
used to measure the alignment between galaxies and structures, such as
filaments and sheets, identified in this cosmic web. We also describe a
similar methodology applied to numerical $N$-body simulations, which we use to
interpret our findings.
\begin{figure*}
\center
\includegraphics[width=0.9\textwidth]{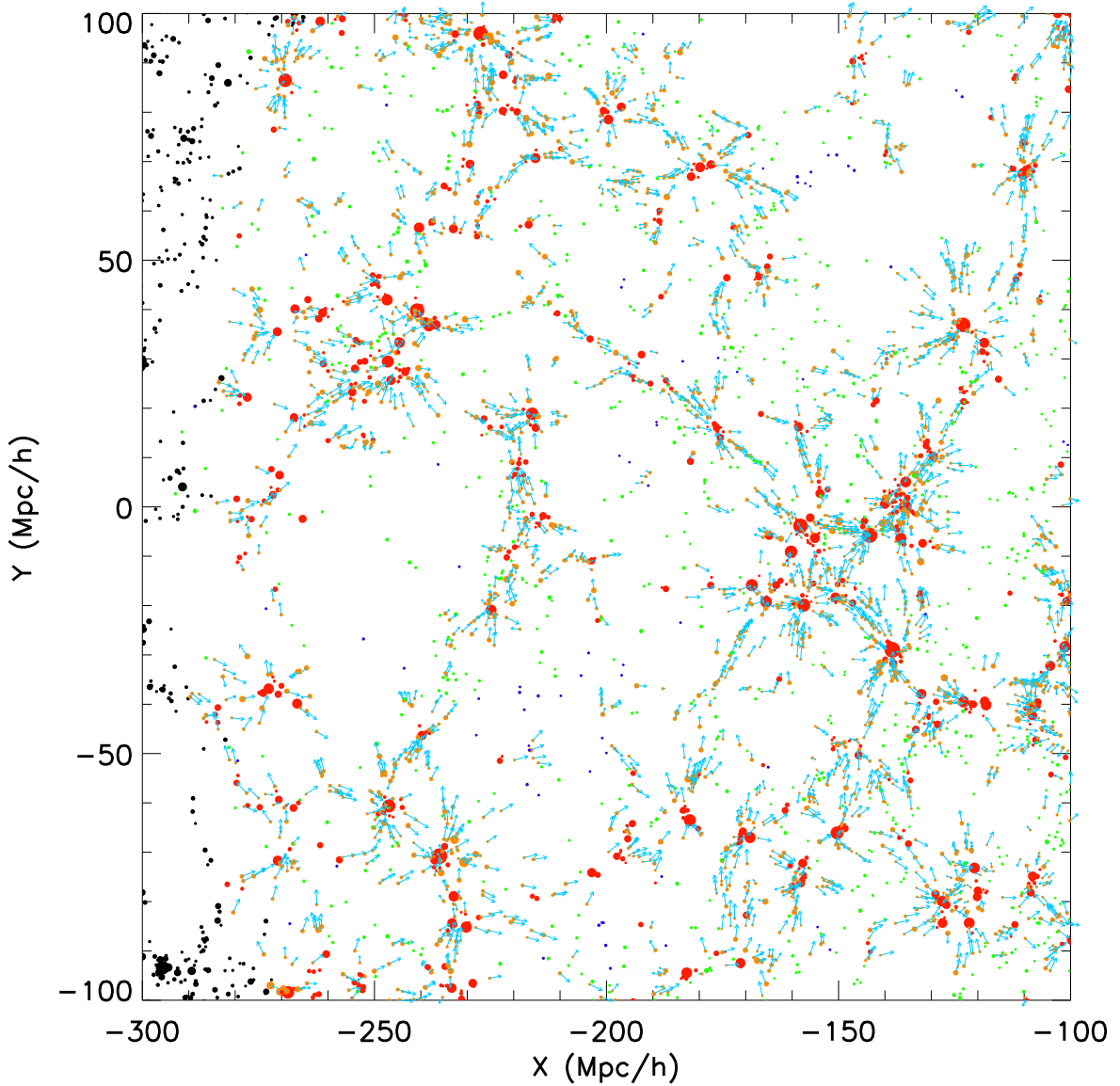}
\caption{Galaxy distributions and environmental classifications in a slice of
  thickness $10\mpc$ from SDSS DR7. The galaxy groups in four different
  environments are classified by different colors: clusters (red), filaments
  (orange), sheets (green),voids (blue).  The cyan arrow indicates the
  direction of the filament at the center of each group. Black dots are groups
  with $z>0.12$ and therefore fall outside our survey volume, i.e., that have
  a filling factor $F = 0$ (see text), and are therefore not included in the
  analysis..  }
\label{fig:filament}
\end{figure*}

\subsection{Observational Data}
\label{sec_SDSS}

The SDSS \citep{York2000}, one of the most influential galaxy redshift
surveys, is a multi-band imaging and spectroscopic survey to explore the
large-scale distribution of galaxies and quasars. The galaxy sample used here
is constructed from the New York University Value-Added Galaxy Catalog
\citep[NYU-VAGC;][]{Bla05}, which is based on the SDSS Data Release 7
\citep{Aba2009}, but with an independent set of significantly improved
reductions over the original pipeline. Marking the completion of the survey
phase known as SDSS-II, DR7 features spectroscopy that is now complete over a
large contiguous area of the Northern Galactic Cap, closing the gap which was
present in previous data releases. The continuity over this large area is a
great advancement and critical to the statistics of large-scale structure.
From the NYU-VAGC, we select all galaxies in the Main Galaxy Sample with an
extinction-corrected apparent magnitude brighter than $r=17.72$, with
redshifts in the range $0.01 \leq z \leq 0.20$ and with redshift completeness
${\cal C}_z > 0.7$. The extracted SDSS galaxy catalog contains a total number
of $639,555$ galaxies with a sky coverage of $\sim 7,750$ square degrees.  We
have complemented this sample with a small subset of galaxies that do not have
SDSS spectra, but for which we have used the redshifts kindly made available
to us from the Korea Institute from Advanced Study (KIAS) Value-Added Galaxy
Catalog \citep{Park2005, Choi2007, Choi2010}.

Galaxy groups are selected using the adaptive halo-based group finder
developed by \citet{Yang2005, Yang2007}. First potential group centers are
identified using the standard FOF algorithm or an isolation criterion. Next,
the total group luminosity, defined as the summed luminosity of all group
members, is converted into an estimate for the group mass using an assumed
mass-to-light ratio. This mass estimate is used to compute the expected
velocity dispersion and angular size of the corresponding dark matter halo,
which in turn are used to select group members in redshift space. This method
is iterated until group memberships converge \citep[see][for
  details]{Yang2007}.

Applying our group finder to the sample of $639,555$ galaxies described above
yields a catalog of $472,532$ galaxy groups (including systems with single
galaxies). Since the group luminosities are strongly correlated with halo
masses, we assign halo masses to our groups based on the ranking of the total,
characteristic group luminosity,  defined as the summed luminosity of all
  group members with $r$-band absolute magnitude $^{0.1}M_r - 5 \log h \leq
  -19.5$.  Here the absolute magnitudes of galaxies are all $K$-corrected and
  evolution corrected to $z=0.1$, using the method described in Blanton et
  al. (2003). The resulting mass assignment is complete to $z \sim 0.12$ for
groups with mass $M_{\rm h} \gtrsim 10^{12} \msun$. Hence, in what follows we
restrict our analysis to the nearby volume covering the redshift range $0.01
\le z \le 0.12 $, and within the large continuous region covering the north
galactic cap. Using all galaxy groups with an assigned mass $M_{\rm h} \ge
M_{\rm th} = 10^{12}\msun$, \citet{Wang2012} constructed the mass, tidal and
velocity (SDSS-MTV) fields for this volume. In what follows, we will use these
fields to quantify the cosmic web. In order to account for boundary effect
that might impact the SDSS-MTV field construction, \citet{Wang2012} introduced
the `filling factor' parameter, $F$, which characterizes the distance of a
galaxy group to the survey boundary. For each group inside the volume covering
the redshift range $0.01 \le z \le 0.12 $, the filling factor $F$ is defined
as the fraction of the survey volume in a sphere of radius $R_F$ centered on
this group. Generally, large $F$ means that the group is located in the inner
region of the survey volume, while small $F$ means the group is near the
boundary. Throughout we adopt $R_F = 80 \mpc$, and we have tested that our
results are insensitive to changes in $R_F$ of a factor of two. We also
emphasize that we have found the alignment signals to be insensitive to how we
cull our sample based on $F$; using only groups with $F > 0.9$ or all groups
with $F > 0$ results in differences in the average misalignment angles (see
below for definition) of $< 0.1^{\circ}$. Hence, in what follow we use all
groups with $F>0$.

\subsection{Characterizing the Cosmic Web}
\label{sec_web}

In specifying the SDSS-MTV fields, \citet{Wang2012} use the tidal tensor field
\begin{equation}\label{TidalTensor}
T_{ij}(\bx) = \frac {\partial^2 \phi} {\partial x_i \partial x_j},
\end{equation}
where $i$ and $j$ are indices with values of $1$, $2$, or $3$, and $\phi$ is
the peculiar gravitational potential, which can be calculated from the
distribution of dark matter halos with mass $M_{\rm h}$ above some threshold
value $M_{\rm th}$ through the Poisson equation,
\begin{equation}\label{PoissonEqn}
\nabla^2 \phi = 4 \,\pi\, G\, \bar{\rho} \, \delta =
4 \, \pi \, G \, \bar{\rho} \, \delta_{\rm h}/b_{\rm h}\,.
\end{equation}
Here $\bar{\rho}$ is the average density of Universe, $\delta(\bx) =
\rho(\bx) /\bar{\rho} - 1$ is the matter overdensity field,
$\delta_{\rm h}(\bx)$ is the overdensity field of dark matter halos
(groups) with mass $M_{\rm h} \ge M_{\rm th}$, whose average, linear
bias parameter is given by $b_{\rm h}$.  Thus, at the location of each
group we can derive $\phi$ and $T_{ij}$ using the distribution of
groups with $M_{\rm h} \ge M_{\rm th}$ \citep[see][]{Wang2012}.
Briefly, we first transform the redshift and sky coordinates
for each group into a Cartesian coordinate system using
\begin{align}\label{coordinate_trans}
X & = R(z) \cos \delta \cos \alpha \nonumber \\
Y & = R(z) \cos \delta \sin \alpha \nonumber \\
Z & = R(z) \sin \delta,
\end{align}
where $\alpha$ and $\delta$ refer to the right ascension and
declination, respectively, and $R(z)$ is the comoving distance out to
redshift $z$. Next, the halo overdensity field, $\delta_{\rm h}$, is
computed on a rectangular $(X,Y,Z)$ grid. This field is then corrected
for redshift space distortions using the method of \citet{Wang2009},
which is based on linear theory. Subsequently, Fast Fourier Transform
is used to obtain the potential field, $\phi$, by solving the Poisson
equation (\ref{PoissonEqn}), after which derivative operators are
applied (in Fourier space) to derive the tidal tensor.

This tidal tensor is subsequently diagonalized to obtain the eigenvalues
$\lambda_1 \geq \lambda_2 \geq \lambda_3$ of the tidal tensor at the position
of the group, which, in analogy with Zel'dovich theory \citep{Zeld70}, can be
used to classify the group's environment in one of four classes:
\begin{itemize}
  \item {\cluster}: a point where all three eigenvalues are positive
  \item {\filament}: a point where $T_{ij}$ has one negative and two
    positive eigenvalues
  \item {\sheet}: a point where $T_{ij}$ has two negative and one
    positive eigenvalues
  \item {\void}: a point where all three eigenvalues are negative
\end{itemize}
(see Hahn et al. 2007a, b).  Of the $277,139$ galaxy groups with
$M_{\rm h} \geq M_{\rm th} =10^{12} \msun$ in our survey volume,
$41,908$ groups ($15.1\%$) are located in a \cluster environment,
$173,820$ groups ($62.7\%$) are located in a \filament, $57,169$ groups
($20.6\%$) are located in a \sheet, and $4,242$ group ($1.5\%$) are
located in a \void. Fig.~\ref{fig:filament} shows the distribution
of galaxy groups from the SDSS DR7 in a $200\mpc \times 200\mpc$ slice
of thickness $10 \mpc$, in which groups in different environment
classes are indicated with different colors \citep[see
  also][]{Wang2012}.

\subsection{Alignment Angles}
\label{sec_angles}

The main goal of this study is to characterize to what extent the orientation
of galaxies is aligned with the structure of its surrounding cosmic web (i.e.,
the orientation of filaments and sheets). To that extent we consider two
alignment angles: for galaxies whose group is located in a \filament
environment, we compute the angle $\theta_{\rm GF}$, defined as the angle
between the orientation of the galaxy, $\theta_{\rm G}$, and the direction of
the filament, $\theta_{\rm F}$.  For galaxies whose group is located in a
\sheet environment, we instead compute the angle $\theta_{\rm GS}$, defined as
the angle between $\theta_{\rm G}$ and the normal direction of the sheet
$\theta_{\rm S}$.

The orientation angle $\theta_{\rm G}$ of the major axis of a galaxy,
projected on the sky, is specified by the $25$ magnitudes per square arcsecond
isophote in the r-band. The orientation angle of a filament, $\theta_{\rm F}$,
at the location $\bx$ of a \filament group is computed as follows. First we
identify the 3D direction $\Delta \bx$ of the filament with the eigenvector
corresponding to the single negative eigenvalue of the tidal tensor at
$\bx$. Next, we compute the {\it projected} direction of the filament on the
sky using
\begin{equation}\label{position_angle}
\theta_{\rm F} = \arctan\left[ \frac{\Delta \alpha \, \cos \delta_{\bx}}
{\Delta \delta}\right],
\end{equation}
where $\Delta \alpha$ and $\Delta \delta$ are the right ascension and
declination differences, respectively, between the locations $\bx$ and $\bx +
\Delta \bx$, and $\delta_{\bx}$ is the declination at $\bx$.  Similarly, the
projected orientation $\theta_{\rm S}$ of the normal vector of a sheet, at the
location $\bx$ of a \sheet group, is computed using the same
equation~(\ref{position_angle}), but now $\Delta \bx$ is the 3D eigenvector
corresponding to the single positive eigenvalue of the tidal tensor at
$\bx$. As an illustration, the cyan arrows in Fig.~\ref{fig:filament} indicate
the orientations of filaments projected along the Cartesian coordinate $Z$ in
equation~(\ref{PoissonEqn}). Note that only groups in a \filament environment,
have such a cyan arrow . Visual inspection of Fig.~\ref{fig:filament} reveals
that the filament directions agree nicely with the prominent filamentary
structures that are visible solely from the spatial distribution of groups in
the cosmic web.

In order to quantify possible alignments and their significance, we generate
$100$ random samples in which the orientations of filaments and sheets
($\theta_{\rm F}$ and $\theta_{\rm S}$) are kept fixed, but the orientations
of the galaxies, $\theta_{\rm G}$, are randomized.  The alignment signal can
then be expressed by
\begin{equation}
P(\theta_{\rm GX}) = {N(\theta_{\rm GX}) \over
\langle N_{\rm R}(\theta_{\rm GX}) \rangle}\,,
\end{equation}
where $X$ either stands for `F' or `S', $N(\theta_{\rm GX})$ is the number
of galaxy-filament or galaxy-sheet pairs in each bin of alignment angle
$\theta_{\rm GX}$, and $\langle N_{\rm R} (\theta_{\rm GX}) \rangle$ is the
average number of such pairs obtained from the $100$ random samples. For the
random samples, we also compute the standard deviation of $P_{\rm
  R}(\theta_{\rm GX}) = {N_{\rm R}(\theta_{\rm GX})}/{\langle N_{\rm R}
  (\theta_{\rm GX}) \rangle}$, which is used to assess the significance of any
detected alignment signal. Since the significance of any alignment is
quantified with respect to the null hypothesis, throughout this paper, we plot
the error bars on top of the $P_{\rm R}(\theta_{\rm GX}) = 1$ line. In
addition to the PDFs, we also calculate the mean angle $\langle \theta_{\rm
  GX} \rangle$ and $\sigma_{\theta_{\rm GX}}$, which is the the standard
deviation of ${\langle \theta_{\rm GX} \rangle}_{\rm R}$ for the $100$ random
samples. The angle $\theta_{\rm GX}$ is constrained in the range $0^\circ \le
\theta_{\rm GX} < 90^\circ$. In the absence of any alignment, $P(\theta_{\rm
  GX}) = 1$ and $\langle\theta_{\rm GX}\rangle =45^\circ$. Values of $\langle
\theta_{\rm GX} \rangle < 45^\circ$ indicate that the galaxy orientations tend
to be parallel to the filaments or the normal directions of the sheets, while
$\langle \theta_{\rm GX} \rangle > 45^\circ$ indicates perpendicular
orientations.

\subsection{Comparison with N-body Simulations}
\label{sec_Nbody}

In order to facilitate the interpretation of our analysis, we also use a
N-body simulation carried out at the Shanghai Supercomputer Center using the
massively parallel TreeSPH GADGET-2 code \citep{Spr2005}. The simulation
evolved $1024^3$ dark matter particles in a periodic box of size $L =
100\mpc$. Glass-like cosmological initial conditions \citep{White1996} were
generated at redshift $z = 120$ using the Zel'dovich approximation. The
particle mass and softening length are $6.93 \times 10^7 \msun$ and $2.25
\kpc$, respectively. The adopted cosmological parameters are $\Omega_{\rm m} =
0.25$, $\Omega_{\rm b} = 0.045$, $\Omega_\Lambda = 0.75$, $h=0.73$, $n=1$ and
$\sigma_8= 0.9$.

Dark matter halos are identified from the $z = 0$ simulation output using the
standard FOF algorithm \citep{Davis1985} with a linking length of 0.2 times
the mean inter-particle separation. To obtain a reliable measurement of the
halo shape, we only retain halos with at least $500$ particles for further
analysis, resulting in a catalog with $73,068$ halos covering the mass range
from $3.5 \times 10^{10} \msun$ to $6.8 \times 10^{14} \msun$.

The shape of a FOF halo containing $N$ particles is determined using
the moment of inertia tensor
\begin{equation} \label{eq:inertia}
I_{\alpha\beta} = m \sum_{i=1}^N x_{i,\alpha} \, x_{i,\beta}\,,
\end{equation}
where $m$ is the mass of a dark matter particle, and $x_{i,\alpha}$
denotes the $\alpha$-component of the position vector of particle $i$
relative to the center of mass. The axis ratios of the halo are
proportional to the square roots of the corresponding eigenvalues, and
the orientation of the halo is determined by the corresponding
eigenvectors.

In the simulation, we have extracted the cosmic filaments and sheets
using the Density Field Hessian Matrix Method described in
\citet{Zhang2009} (see also Hahn et al. 2007b). This method uses the
eigenvalues of the Hessian matrix of the smoothed mass density field
(smoothed with a Gaussian filter with a smoothing length $R_{\rm s} =
2.1 \mpc$) to classify the cosmic web environment of a dark matter
halo, similar to the method used above based on the tidal tensor.  The
directions of the filaments and sheets are again based on the
directions of the eigenvectors as in \S\ref{sec_angles}.  In
\citet{Zhang2009}, this simulation and methodology were used to study
the alignment between the spins and shapes of dark matter halos and
their large scale environment. They find that both the spin and the
major axes of \filament halos with masses $\lesssim 10^{13} \msun$ are
preferentially aligned with the direction of the filaments. The spins
and major axes of \sheet halos were found to lie preferentially
parallel to the sheets \citep[see also][]{Altay2006,Arag2007,Hahn2007a}.

Here, in order to compare with the alignment signals detected in the
SDSS data, we project the orientations of the halos, filaments and
sheets along one of the simulation's coordinate axes. Subsequently, we
compute the probability distribution functions $P(\theta_{\rm HF})$
and $P(\theta_{\rm HS})$, where $\theta_{\rm HF}$ is the angle between
the projected orientation of the major axis of a \filament halo and
that of its filament, while $\theta_{\rm HS}$ is the angle between the
projected orientation of the major axis of a \sheet halo and the
projected normal axis of the sheet.

\section{Results}
\label{sec_result}

To quantify the impact of filaments and sheets onto galaxies, we investigate
the orientations of galaxies relative to the directions of the filaments or
the normal vectors of the sheets in which the galaxies reside. For simplicity,
we will often use the terms ``shape'', as well as ``filament'' and ``sheet'',
as an indication of the direction in the obvious sense.

\subsection{The Alignment between Galaxy Shape and Filament}
\label{sec_gf}

\begin{figure}
\center
\includegraphics[width=0.4\textwidth,height=0.35\textwidth]{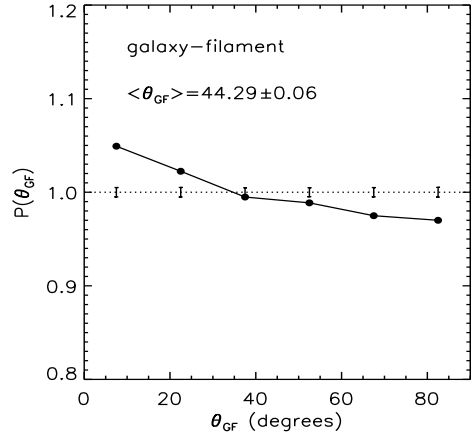}
\caption{Normalized probability distribution of the angle $\theta_{\rm
    GF}$ between the projected orientation of the major axis of SDSS
  galaxies and that of the filament in which its group resides. The
  horizontal dotted line corresponds to an isotropic distribution of
  alignment angles, while the error bars indicate the scatter obtained
  from $100$ realizations in which the orientations of the galaxies
  have been randomized. The average value of $\theta_{\rm GF}$ and its
  error (obtained from the $100$ random realizations) are indicated.}
\label{fig:gf}
\end{figure}
\begin{figure}
\center
\includegraphics[width=0.4\textwidth,height=0.35\textwidth]{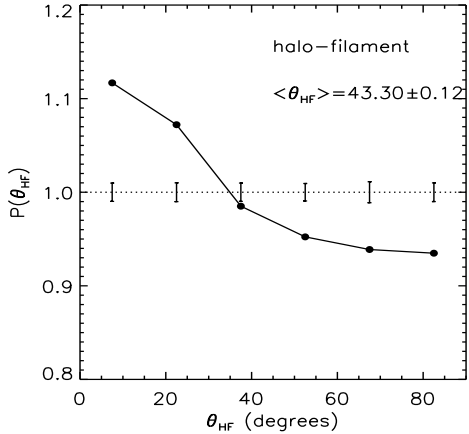}
\caption{Same as Fig.~\ref{fig:gf}, but for the alignment angle
  $\theta_{\rm HF}$ between the projected orientation of a dark matter
  halo and that of the filament in which it resides. These results
  have been obtained using the numerical $N$-body simulation described
  in \S\ref{sec_Nbody}.}
\label{fig:gf_sim}
\end{figure}

Fig.~\ref{fig:gf} shows the probability distribution $P(\theta_{\rm
  GF})$ of the angle between the projected major axis of \filament
galaxies and the projected direction of their filaments, as measured
from the SDSS DR7 galaxy catalog using the method described in
\S\ref{sec_data}. With an average alignment angle of $\langle
\theta_{\rm GF} \rangle = 44.29^{\circ} \pm 0.06^{\circ}$, there is a
clear, and highly significant, indication that \filament galaxies are
preferentially aligned parallel to their filaments

Fig.~\ref{fig:gf_sim} shows the equivalent plot for $\theta_{\rm HF}$, the
alignment angle between the projected orientation of \filament dark matter
halos and that of the filament in which they reside.  As in the SDSS data,
their is a clear and significant alignment between halos and filaments, in the
sense that the major axis of dark matter halos preferentially aligns parallel
to the filament in which it resides. The average alignment angle of $\langle
\theta_{\rm HF} \rangle = 43.30^{\circ} \pm 0.12^{\circ}$ is smaller than for
the SDSS galaxies.   In addition, from N-body simulation we select a
  sample of dark matter halos, with mass distribution equivalent to the host
  halos of central galaxies in SDSS galaxy catalog. We find the alignment
  signals for the dark matter halos are stonger than those of central galaxies
  in SDSS galaxy catalog, and the average alignment angle is 
$\langle \theta_{\rm HF} \rangle = 42.19^{\circ} \pm 0.38^{\circ}$.  This
suggests a net misalignment between the major axes of galaxies and their dark
matter halos, in excellent agreement with a number of previous alignment
studies \citep[e.g.,][]{Kang2007, WangY2008, Oku2009, Agu2010, LiC2013}.

\subsubsection{Dependence on Galaxy Properties}
\label{sec_gfprop}

\begin{figure}
\center
\includegraphics[width=0.48\textwidth,height=0.4\textwidth]{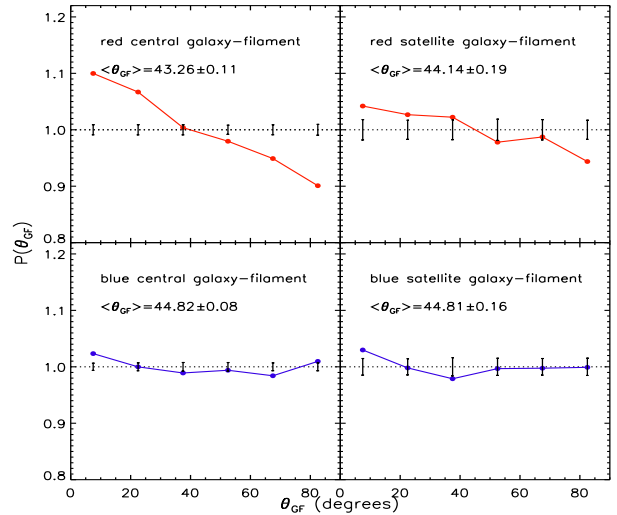}
\caption{Same as Fig.~\ref{fig:gf}, but for different subsamples
  (`red' and `blue') of central and satellite galaxies, as indicated.}
\label{fig:gf_rc}
\end{figure}

The large number ($212,046$) of \filament galaxies in our sample
allows us to study how the galaxy-filament alignment signal depends on
various galaxy properties. We start by using the galaxy group catalog
to split the galaxy population into centrals (defined as the brightest
group members) and satellites (those group members that are not
centrals).  These are further subdivided in `red' and `blue' according
to their $^{0.1}(g-r)$ colors: galaxies with $^{0.1}(g-r) \geq 0.83$
are called `red', while those with $^{0.1}(g-r) < 0.83$ are called
`blue'. The value $0.83$ roughly corresponds to the bimodal scale in
the color-magnitude relation \citep{Weinmann2006}.

Fig.~\ref{fig:gf_rc} shows the galaxy shape-filament alignment signals
for the resulting four sub-samples. Central galaxies reveal a
remarkably strong color-dependence. While red centrals show strong
alignment along their filaments, with $\langle \theta_{\rm GF} \rangle
= 43.26^{\circ} \pm 0.11^{\circ}$, virtually identical to that of the
dark matter halos (cf. Fig.~\ref{fig:gf_sim}), blue \filament centrals
are only marginally aligned, with an average misalignment angle that
is consistent with no alignment at the $2.3\sigma$-level.  Satellite
galaxies, overall, are less strongly aligned with the filaments in
which their host-groups reside than centrals. In particular, the
orientations of blue satellites are consistent at the
$1.2\sigma$-level with having a random (projected) orientation with
respect to their filament. In the case of red satellites, however, the
alignment signal is $\langle \theta_{\rm GF} \rangle = 44.14^{\circ}
\pm 0.19^{\circ}$, which is significant at the $4.5\sigma$-level.

A quantitatively similar dependence on galaxy color has been found for
the alignment between the orientation of central galaxies and the
angular distribution of their satellites \citep{Yang2006, Azz2007,
  WangY2008, Agu2010}, and suggests that red centrals are more
accurately aligned with their host halo than blue centrals. We
caution, though, that, as demonstrated by \citet{Kang2007}, the
interloper fraction (in the group catalog) is larger among blue
centrals than among red centrals, which may cause a stronger dilution
of the alignment signal for the blue centrals (since satellite
galaxies show a weaker alignment signal). We will return to the
interpretation of these findings in \S\ref{sec_summary}.

\begin{figure}
\center
\includegraphics[width=0.49\textwidth]{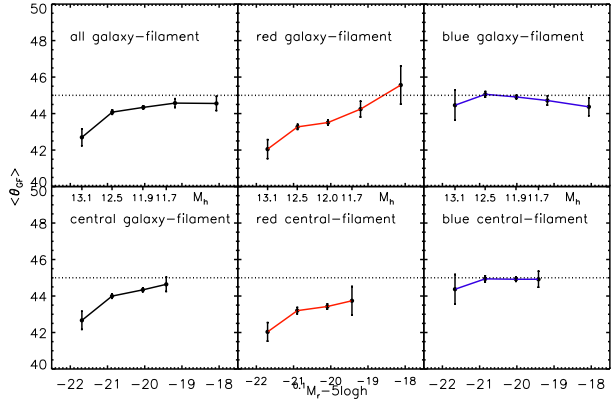}
\caption{Average angle $\langle\theta_{\rm GF}\rangle$ as a function
  of the absolute magnitude $^{0.1}M_r - 5 \log h$ for \filament 
  galaxies. The average halo masses that correspond to given
  r-band magnitudes of central galaxies are indicated in
  the top axes of the lower panels.}
\label{fig:gf_mmh}
\end{figure}

Finally, we also investigate how the galaxy-filament alignment strength
depends on the luminosity (stellar mass) of the galaxy, as well as on the mass
of its dark matter host halo. To that extent we split the galaxy sample in
equal sub-samples according to their $r$-band magnitude, and calculate
$\langle\theta_{\rm GF}\rangle$ for each subsamples. Fig.~\ref{fig:gf_mmh}
shows $\langle\theta_{\rm GF}\rangle$ versus $^{0.1}M_r - 5 \log h$.  Results
are shown separately for all color (left panels), red (middle panels) and blue
galaxies (right panels), as well as separately for all (upper panels) and
central (lower panels) galaxies. As we have the halo mass estimation for
  each group, we can obtain the average halo masses that correspond to given
  $r$-band magnitudes of central galaxies, which are indicated in the top axes
  of the lower raw panels. Overall, the alignment signal is stronger for
brighter (massive) and redder galaxies, and for centrals in more massive
halos.  For blue galaxies the alignment is very weak at luminous end and
almost absent at faint end. For red galaxies, significant alignment exist for
all galaxies brighter than $^{0.1} M_r- 5\log h = -20$, corresponding to a
halo mass of $10^{12}h^{-1}{\rm M}_\odot$ for central galaxies.

A similar dependence of the halo-filament alignment on halo mass has been
found in our N-body simulations.  More massive halos in filaments have
stronger alignment signals \citep{Zhang2009}. Here we perform the similar
  analysis for halos in difference mass ranges, we find $\langle\theta_{\rm
    HF}\rangle = 42.18\pm0.61$ at $\log(M_h)=12.0$, $\langle\theta_{\rm
    HF}\rangle = 41.25\pm1.05$ at $\log(M_h)=12.5$. Again the alignment
  signial is slightly (significantly) larger than that of red (blue) central
  galaxies in the same mass range.

\subsection{The Alignment between Galaxy Shape and Sheet}
\label{sec_gs}

\begin{figure}
\center
\includegraphics[width=0.4\textwidth,height=0.35\textwidth]{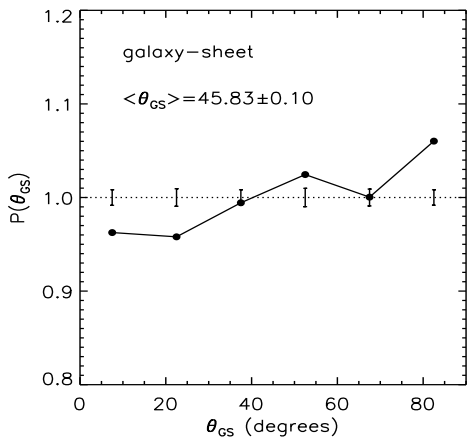}
\caption{Normalized probability distribution of the angle $\theta_{\rm GS}$
  between the projected orientation of the major axis of SDSS galaxies and
  that of the normal vector of the sheet in which its group resides. As
  before, the error bars indicate the scatter obtained from $100$ realizations
  in which the orientations of the galaxies have been randomized.}
\label{fig:gs}
\end{figure}

\begin{figure}
\center
\includegraphics[width=0.4\textwidth,height=0.35\textwidth]{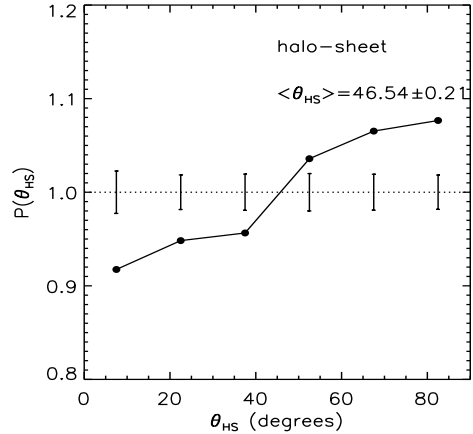}
\caption{Same as Fig.~\ref{fig:gs}, but for the alignment angle $\theta_{\rm
    HS}$ between the projected orientation of a dark matter halo and that of
  the sheet in which it resides. These results have been obtained using the
  numerical $N$-body simulation described in \S\ref{sec_Nbody}.}
\label{fig:gs_sim}
\end{figure}

Having discussed the alignment between galaxies and filaments, we now turn our
attention to the galaxy-sheet alignment. To that extent we now focus on the
$61,662$ member galaxies of the $57,169$ groups that reside in a \sheet
environment, and measure the probability distribution functions for their
alignment angle $\theta_{\rm GS}$, between the projected orientations of the
galaxy and the normal vector of the sheet.

The results for all \sheet galaxies are shown in Fig.~\ref{fig:gs}, which
indicate a significant anti-alignment between galaxy major axis and
sheet-normal of $\langle \theta_{\rm GS} \rangle = 45.83^\circ \pm
0.10^\circ$. This indicates that \sheet galaxies have their minor axes
preferentially aligned along the normal vector of the sheet.
Fig.~\ref{fig:gs_sim} shows the alignment signal between the projected
orientation of dark matter halos and the normal vector of their sheets, as
obtained from the $N$-body simulation discussed in \S\ref{sec_Nbody}. Similar
to the SDSS data, their is a significant anti-alignment.  Although the mean
misalignment angle ($46.54^{\circ}$) is somewhat larger than that for the
galaxies, this difference is, given the larger errorbar for the dark matter
halos, not significant. These results regarding the dark matter halos are in
qualitative agreement with a number of previous studies based on numerical
simulations in $\Lambda$CDM cosmologies \citep[e.g.,][] {Arag2007, Hahn2007b,
  Zhang2009}.

\subsubsection{Dependence on Galaxy Properties}
\label{sec_gsprop}

\begin{figure}
\center
\includegraphics[width=0.48\textwidth,height=0.4\textwidth]{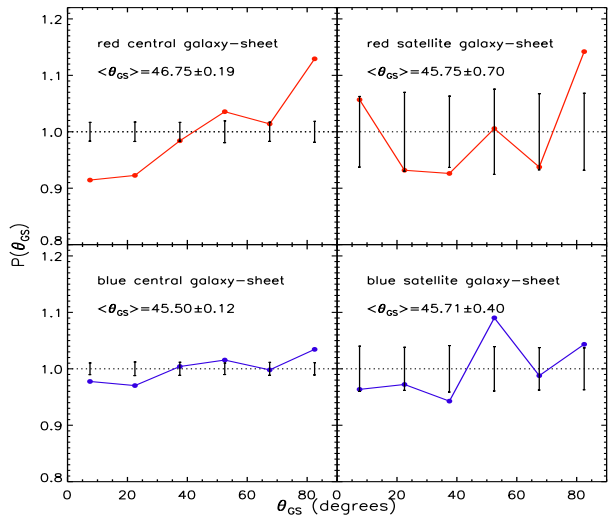}
\caption{ Same as Fig.~\ref{fig:gf_rc} but for the galaxy-sheet
  alignment angle, $\theta_{\rm GS}$.  }
\label{fig:gs_rc}
\end{figure}

In the following, we investigate the dependence of the galaxy-sheet alignment
on the galaxy properties.  Similar to what we did for the galaxy-filament
alignments, we first split the population in centrals and satellites, which
are then further subdivided in `red' and `blue', based on their $^{0.1}(g-r)$
color. As shown in Fig.~\ref{fig:gs_rc}, the color dependence of the
shape-sheet alignment is similar to that of the shape-filament alignment. Red
galaxies are more strongly aligned with the plane of the sheet than blue
galaxies, and the alignment for centrals is much stronger than for
satellites. In fact, neither red nor blue satellites show a significant (more
than $2\sigma$) alignment signal. The strongest alignment signal is found for
red centrals, for which $\langle \theta_{\rm GS} \rangle = 46.75^\circ \pm
0.19^\circ$ (a $9.2\sigma$ significant detection of alignment). Interestingly,
the strength of this alignment signal is again very similar to that of the
dark matter halos in the simulation (cf. Fig.~\ref{fig:gs_sim}), indicating
once more that red centrals may be strongly aligned with their host halos.
\begin{figure}
\center
\includegraphics[width=0.49\textwidth]{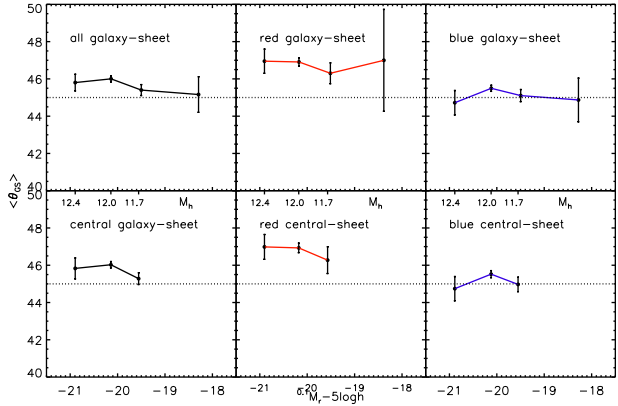}

\caption{ Same as Fig.~\ref{fig:gf_mmh} but for the galaxy-sheet
  alignment angle, $\langle \theta_{\rm GS} \rangle$.}
\label{fig:gs_mmh}
\end{figure}

Finally, Fig.~\ref{fig:gs_mmh} shows the dependence of $\langle \theta_{\rm
  GS}\rangle$ on galaxy luminosity for subsamples divided according to color,
and according to central versus total.  In agreement with the results for the
galaxy-filament alignment, the alignment is typically stronger and significant
only for red galaxies.  Within a given color subsample, the dependence on
galaxy luminosity is rather weak.

\subsection{The Alignment between Galaxy Shape and Tidal Tensor}
\label{sec_gt}

In the previous subsections, we investigated the alignments of \filament and
\sheet galaxies with their respective cosmic web structure. This analysis,
however, ignored the \cluster and \void galaxies, which combined make up about
30\% of the galaxy population in our sample. In this subsection, we test the
alignment of the {\it entire} galaxy population ($395,841$ galaxies in
\cluster, \sheet, \filament, and \void environments) with the local tidal
field.  As discussed in \S\ref{sec_web}, the latter has been calculated from
the SDSS DR7 using the distribution of galaxy groups with assigned halo masses
$M_{\rm h} \geq 10^{12} \msun$. At the center of mass of each group we
calculate the eigenvalues $\lambda_1$, $\lambda_2$ and $\lambda_3$ ($\lambda_1
> \lambda_2 > \lambda_3$) of the tidal tensor. In what follows we refer to the
corresponding eigenvectors as $\bt_1$, $\bt_2$, and $\bt_3$, respectively. The
projected orientation of $\bt_1$, $\bt_2$, and $\bt_3$, at the location $\bx$
of a group, is computed using the equation~(\ref{position_angle}), where
$\Delta \bx$ is the 3D eigenvector corresponding to the eigenvalue
$\lambda_1$, $\lambda_2$ and $\lambda_3$, respectively.
\begin{figure*}
\center
\includegraphics[width=0.9\textwidth]{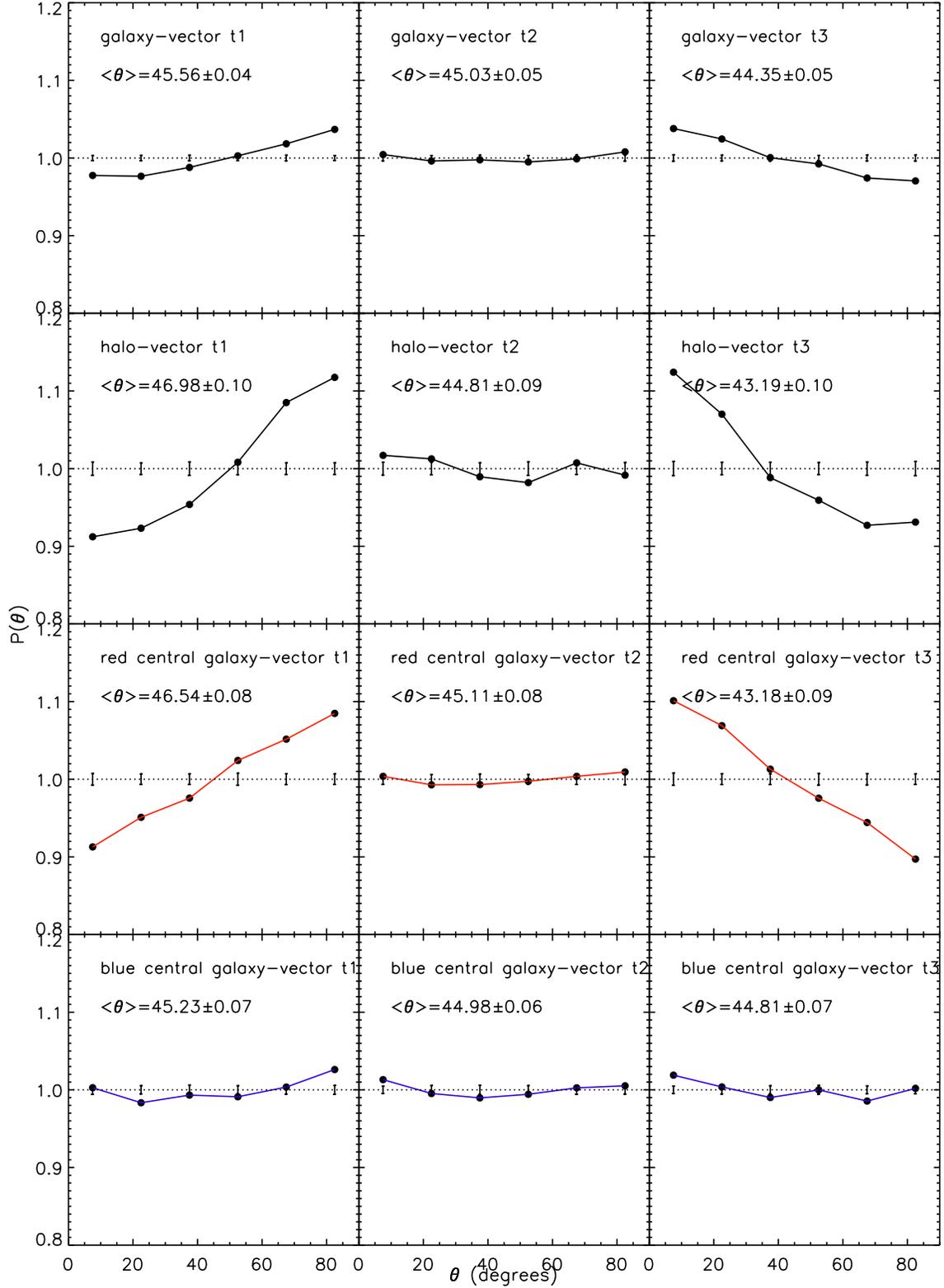}
\caption{Probability distribution of the angle $\theta$ between the
  orientation of the (projected) major axis of the galaxy and the
  eigenvectors $\bt_1$ (left column), $\bt_2$ (middle column), and
  $\bt_3$ (right column) of the tidal tensor field for all galaxies in
  the SDSS (first row), for dark matter halos in our $N$-body
  simulation (second row), for red central galaxies in the SDSS (third
  row), and for blue central galaxies in the SDSS (fourth row). The
  eigenvector $\bt_i$ corresponds to the
  eigenvalue $\lambda_i$, where $\lambda_1 \geq \lambda_2 \geq
  \lambda_3$. As before, errorbars indicate the scatter obtained from
  $100$ realizations in which the orientations of the galaxies or dark
  matter halos have been randomized.}
\label{fig:tf}
\end{figure*}

The first row of Fig.~\ref{fig:tf} shows the alignment between the projected
orientations of the three eigenvectors of the tidal field and that of the
major axis of the galaxies. For galaxies in \sheet environments, $\bt_1$
indicates the direction perpendicular to the sheet-plane, and the alignment
with $\bt_1$ is therefore identical to $\theta_{\rm GS}$. Similarly, the the
alignment with $\bt_3$ is identical to $\theta_{\rm GF}$ for \filament
galaxies. Here, though, we include all galaxies, independent of their
environment, and we show the alignment angles with all three eigenvectors of
the tidal tensor. There is a very significant alignment ($14\sigma$) with the
eigenvector associated with the largest eigenvalue, $\bt_1$, which is expected
from the fact that we had already seen (\S\ref{sec_gs}) that \sheet galaxies
are preferentially oriented perpendicular to the normal vector of the
sheet. There is no indication of any alignment between galaxies and the
intermediate eigenvector, $\bt_2$, but again a very significant alignment
($13\sigma$) with the eigenvector associated with the smallest eigenvalue,
$\bt_3$. The latter is again expected from the fact that \filament galaxies
(which make up 53.6\% of the {\it entire} galaxy population) are aligned with
the direction of their filaments.

The second row in Fig~\ref{fig:tf} shows the alignments between the projected
orientations of dark matter halos and the eigenvectors of the tidal tensor
(also in projection), obtained from the $N$-body simulation. Halos are
anti-aligned with $\bt_1$, positively aligned with $\bt_3$, and not aligned
with $\bt_2$, in excellent agreement with previous studies
\citep[e.g.,][]{Wang2011, Lib2013, Tempel2013a}.  These results are also in
good, qualitative agreement with the alignments seen for the SDSS galaxies,
although the alignment for halos is found to be significantly stronger than
that for the entire galaxy population. As shown above, red central galaxies
seem to be far more strongly aligned with their sheets or filaments than blue
centrals or satellites. The third row of Fig~\ref{fig:tf} therefore shows the
alignments of red centrals with the eigenvectors of the tidal tensor. These
are found to be amazingly similar to those of the dark matter halos,
supporting the notion that red centrals may be strongly aligned with their
host halos. In contrast, blue centrals do not show any significant alignments
with the eigenvectors of the tidal tensor (the fourth row of the figure). 

\section{Summary and Discussion}
\label{sec_summary}

Using galaxy groups selected from the SDSS DR7, we have examined the alignment
between the major axes of galaxies and their surrounding large-scale
structure, especially the filaments and sheets.  We characterized the large
scale structure using the tidal tensor, which we reconstructed using galaxy
groups with an assigned halo mass $M_{\rm h} \gtrsim 10^{12} \msun$. Based on
the eigenvalues of the tidal tensor at the location of each galaxy group, we
have split the galaxy population into four environments: \cluster, \filament,
\sheet and \void.

We have shown that the major axes of galaxies in a \filament environment are
preferentially aligned along the direction of the filament (defined as the
direction of the eigenvector of the tidal tensor associated with the single
negative eigenvalue), while galaxies in a \sheet environment are
preferentially aligned perpendicular to the normal vector of the sheet
(defined as the direction of the eigenvector of the tidal tensor associated
with the single positive eigenvalue). In both cases, the alignment is
strongest for red central galaxies, is weak for blue centrals, and is
virtually absent for satellite galaxies\footnote{With the exception of a
  $4.5\sigma$ significant alignment for red \filament satellites with the
  direction of their filament.}. In addition, the alignment strength is found
to increase with galaxy luminosity and/or stellar mass.

We also computed the alignments between the orientations of {\it all} galaxies
(independent of their environment) and the three eigenvectors of the tidal
tensor at the location of the galaxy group in which they reside. Consistent
with the above results, we find a strong anti-alignment (at $14\sigma$
significance) with the eigenvector $\bt_1$ associated with the largest
eigenvalue, no alignment at all with the eigenvector $\bt_2$ associated with
the intermediate eigenvalue, and strong, positive alignment (at $13\sigma$
significance) with the eigenvector $\bt_3$ assocated with the smallest
eigenvalue. In the case of red, central galaxies, the significance of these
alignments becomes even stronger: $19.3\sigma$ and $20.2\sigma$ in the cases
of the alignments with $\bt_1$ and $\bt_3$, respectively.  In contrast, for
blue galaxies the alignments are always quite weak.

Using a large, $N$-body simulation, we also calculated how the shapes of dark
matter halos are aligned with their filaments and sheets (in projection). Our
results are in excellent agreement with previous studies
\citep[e.g.,][]{Hahn2007b, Zhang2009, Wang2011, Lib2013,
  Tempel2013a}. Interestingly, we find alignment strengths for the dark matter
halos that are indistinguishable from those of red centrals, strongly
suggesting that red central galaxies are strongly aligned with their dark
matter host halos. This is consistent with the results obtained in
\citet{Yang2006, Kang2007, WangY2008, Oku2009}, all of which show that red
centrals and their host halos are more strongly aligned than the halo -
filament and halo - sheet alignments found here.   These measured
  alignments between galaxy shape and the surrounding large-scale structure
  from observation may indicate that galaxy formation is affected by
  large-scale environments. The alignment signals may be a natural result of
how galaxies accrete material via streams along the direction of the
filaments, and the thin plane of the sheets. They also suggest that the shapes
and orientations of red central galaxies are significantly affected by such
accretion.

The fact that blue centrals show an alignment signal that is much weaker than
that of dark matter halos implies that blue central galaxies are only poorly
aligned with the shape of their host halos.  These results are in qualitative
agreement with previous studies \citep[e.g.,][]{Yang2006, Azz2007, WangY2008,
  Fal2009, Oku2009, Agu2010, LiC2013}, most of which focussed on the alignment
of central galaxies with the spatial (angular) distribution of satellite
galaxies.  The fact that the projected orientation of blue centrals is
positively aligned with that of the filaments (albeit at a marginaly level),
is also interesting in light of some recent results by \citet{Tempel2013b} and
\citet{Tempel2013c}, who found evidence that the rotation axes of disk
galaxies are (weakly) aligned with the directions of their filaments.  A
similar result was obtained by \citet{Hahn2010}, albeit for {\it massive} disk
galaxies in hydrodynamical simulations. Using $N$-body simulations,
\citet{Wang2011} found that the spin axis of a dark matter halo is correlated
with the directions of the tidal directions, but the correlation is
significantly halo mass dependent.  For halos with masses as low as $\sim
10^{12}h^{-1}{\rm M}_\odot$, the correlation becomes rather weak but is still
significant.  In as far as `blue' galaxies are `disk' galaxies with spin axis
similar to their host halos, our results are not consistent with these
results. We caution, though, that all three studies performed their analysis
in 3D, while our analysis is done in projection in 2D.  In addition, all
studies only found a positive alignment between spin vector and filament
direction for {\it massive} disk galaxies and halos.  It might well be that
projection effects, and the inclusion of large numbers of less massive
galaxies, largely erase the signatures of the mild alignment effects seen in
\citet{Hahn2010}, \citet{Wang2011}, \citet{Tempel2013b}, and
\citet{Tempel2013c}.

If galaxies are strongly aligned with sheets and filaments, as we find here at
least for red central galaxies, then galaxies should also be mutually aligned
over scales of those sheets and filaments. Such alignments have important
ramifications for cosmic shear measurements based on the orientation
correlation of source galaxies \citep[e.g.][]{Oku2009, Kirk2012,
  Joachimi2013}.  Our results can help in modeling such contaminations.  In
particular, our finding that the alignments are strong only for red bright
galaxies suggests that the contaminations can be reduced by using only blue
and relatively faint sources.

\section*{Acknowledgements}

This work is supported by the national science foundation of China (grant
Nos. 11203054, 10925314, 11128306, 11121062, 11233005, 11073017),
NCET-11-0879, the CAS/SAFEA International Partnership Program for Creative
Research Teams (KJCX2-YW-T23), and the Shanghai Committee of Science and
Technology, China (grant No. 12ZR1452800).  HJM would like to acknowledge the
support of NSF AST-0908334 and NSF AST-1109354.

Funding for the SDSS and SDSS-II was provided by the Alfred P. Sloan
Foundation, the Participating Institutions, the National Science Foundation,
the U.S. Department of Energy, the National Aeronautics and Space
Administration, the Japanese Monbukagakusho, the Max Planck Society, and the
Higher Education Funding Council for England. The SDSS was managed by the
Astrophysical Research Consortium for the Participating Institutions.

\end{document}